\documentclass[prd,superscriptaddress,nofootinbib,notitlepage]{revtex4-1}

\usepackage{slashed}
\usepackage{caption}
\usepackage{subfigure}
\usepackage{amsmath,amssymb,graphicx,xcolor,mathtools}
\usepackage[colorlinks=true,linktocpage=green,linkcolor=blue,citecolor=blue]{hyperref}
\numberwithin{equation}{section}
\allowdisplaybreaks
\usepackage{hyperref}
\hypersetup{
	colorlinks=true,
	citecolor = blue,
	linkcolor= magenta,
	filecolor=magenta,      
	urlcolor=blue,
	pdftitle={Bulk Viscosity-Rotation}
}
\usepackage{graphics}
\usepackage{mathrsfs}
\usepackage{simpler-wick}

\DeclareMathOperator{\Tr}{Tr}

\def\be {\begin{equation}}
	\def\ee {\end{equation}}
\def\bea {\begin{eqnarray}}
	\def\eea {\end{eqnarray}}
\def\bc {\begin{center}}
	\def\ec {\end{center}}
\def\nn {\nonumber}

\def\wtg{\widetilde{\gamma}}
\def\wtD{\widetilde{D}}
\def\ovpsi{\overline{\psi}}

\DeclareMathAlphabet{\mathpzc}{OT1}{pzc}{m}{it}

\def\sumintof{\sum\!\!\!\!\!\!\!\!\!\int\limits_{\{P\}}}

\begin{document}
	
	\title{Bulk viscosity of rotating, hot and dense spin 1/2 fermionic systems from correlation functions}

	\author{Sarthak Satapathy}
	\email{sarthaks680@gmail.com}
	\affiliation{Department of Physics, Indian Institute of Technology Roorkee, Roorkee - 247667, India
	}

\begin{abstract}

In this work we have presented the one-loop calculation of the bulk viscosity of a system of rotating, hot and dense spin 1/2 fermions within the framework of Kubo formalism calculated from correlation functions of fields which in turn is used to calculate the spectral function of energy-momentum tensors. The calculation has been done in curved space by the help of tetrad formalism, where the the gamma matrices in this set-up assume their generic structure by becoming space dependent. The techniques of thermal field theory have been employed which take into account the three energy scales viz. temperature, chemical potential and angular velocity into account in the Matsubara frequency summation. The study has been performed in the ambience of very large angular velocities, ranging from 0.1 to 1.0 GeV. The fermion propagator used in this work is appropiate for the regime of large angular velocities. We explore the behaviour of bulk viscosity with angular velocity, temperature and chemical potential through our plots.

\end{abstract}
	
\maketitle
	
\section{Introduction}
\label{sec:intro}

The study of Quark Gluon Plasma (QGP) created in heavy-ion collisions \cite{Muller,Heinz:2000bk,Sarkar:2010zza} has gained a lot of popularity when subjected to extreme environments such as extremely high magnetic fields \cite{Kharzeev:2012ph} and large angular velocities \cite{Becattini:2021lfq,Bhadury:2021oat}. Angular velocities of very high magnitude are produced in non-central heavy-ion collisions in the early stages of the formation of QGP as reported by STAR collaboration \cite{STAR:2017ckg}, where the angular momentum is estimated to be of the order $10^{3\sim 5}\hbar$ \cite{Becattini:2021lfq}. In Ref. \cite{STAR:2017ckg} the first measurement of spin polarization i.e the alignment between the angular momentum of a non-central collision and the spin of emitted particles, for $\Lambda$ hyperons was performed showing that this is the most vortical system \cite{Becattini:2016gvu} ever produced in heavy-ion collisions. The particles emitted from such a highly vortical system \cite{Becattini:2016gvu} are expected to be spin polarized \cite{Becattini:2016gvu} in the direction of their angular momentum which have been explored employing different approaches \cite{Son:2009tf,Kharzeev:2010gr,Florkowski:2017ruc,Florkowski:2017dyn,Florkowski:2018ahw,Florkowski:2018fap,Singh:2020rht,Singh:2021man,Florkowski:2021wvk,Bhadury:2020puc,Bhadury:2020cop,Hattori:2019lfp,Fukushima:2020ucl,Li:2020eon,Montenegro:2020paq,Weickgenannt:2020aaf,Garbiso:2020puw,Gallegos:2021bzp,Gallegos:2020otk,Gallegos:2022jow,Sheng:2021kfc,Speranza:2020ilk,Becattini:2016gvu,Karpenko:2016jyx,Pang:2016igs,Xie:2017upb,Becattini:2017gcx,Fu:2020oxj,Florkowski:2021xvy,Montenegro:2018bcf,Serenone:2021zef,Montenegro:2022dth,She:2021lhe,Daher:2022xon,Cao:2022aku,Hu:2021lnx,Hidaka:2018ekt,Yang:2020hri,Wang:2020pej,Yan:2023gst,Erdmenger:2022nhz,Weickgenannt:2021cuo,Hu:2021pwh,Hu:2022lpi,Das:2022azr,Weickgenannt:2022zxs,Stephanov:2012ki,Chen:2014cla,Gorbar:2017toh,Shi:2020htn,Heller:2020hnq,Gallegos:2020otk,Hongo:2021ona,Gallegos:2022jow,Singh:2022uyy,Singh:2022ltu}. Some notable work that have explored the dynamics of spin polarization has been studied in variety of scenarios are entropy-current analysis \cite{Hattori:2019lfp}, Gubser-expanding background \cite{Singh:2020rht}, vorticity and non-local collisions \cite{Weickgenannt:2020aaf}, boost-invariant hydrodynamical background \cite{Florkowski:2019qdp} and an effective spacetime dependent mass approach \cite{Bhadury:2023vjx}. A brief review of relativistic hydrodynamics with spin has been done in 
Ref. \cite{Bhadury:2021oat}, where spin kinetic coefficients have been calculated and the results have been used to calculate the spin polarization of $\Lambda$ hyperons. A very detailed study on the collective dynamics of spin-1/2 fermions can be found in 
Ref. \cite{Singh:2022uyy}. A brief review on local and global spin polarization has been covered in Ref. \cite{Wang:2017jpl}. From a holographic perspective, spin hydrodynamics has been studied in Ref. \cite{Gallegos:2022jow,Gallegos:2020otk,Yan:2023gst,Erdmenger:2022nhz}. A recent review of spin hydrodynamics from a holographic perspective, highliting the prospects and limitations has been discussed in Ref. \cite{Amano:2023bhg}. From a quantum field theoretical approach \cite{Srednicki:2007qs,Schwartz:2014sze}, spin hydrodynamics for spin 1/2 fermions has been studied in a torsionful background in Ref. \cite{Hongo:2021ona}, where field theoretical techniques have been employed for the calculation of transport coefficients of a spinful relativistic fluid. In Ref. \cite{Hongo:2021ona}, particularly the extra transport coefficients corresponding to the antisymmetric part of the energy-momentum tensor \cite{Hehl:1976vr,Becattini:2011ev,Becattini:2012pp,Weickgenannt:2022jes,Speranza:2020ilk,Dey:2023hft,Becattini:2014dsa}, namely rotational viscosity \cite{Hu:2021lnx} has been discussed. Since magnetic field and vorticity both are produced in off-central heavy-ion collisions, a seminal work concerning this topic by Bhadury {\it et.al} \cite{Bhadury:2022ulr} i.e spin magnetohydrodynamics, has been done using the kinetic theory. Recent work by Kiamari {\it et.al} \cite{Kiamari:2023fbe}, have generalized a relativistic spinful and vortical fluid to relativistic magnetohydrodynamics (MHD), which they call as the Spinful-Vortical MHD (SVMHD). Some of the works concerned in this direction have been carried out in Ref. \cite{Singh:2022ltu,Buzzegoli:2022qrr,Peng:2022cya}, where the time evolution of the spin polarization tensor has been studied by solving the spin-MHD equations. Notable field theoretical calculations that have taken into account, the effect of magnetic field and rotation can be found from Ref. \cite{Das:2023vgm}, where the thermal di-lepton production rate (DPR) has been calculated for a system of hot, dense, rotating and magnetized QGP. The results show a suppression of DPR when compared with the non-rotating case in strong magnetic field.\\     
Theoretical studies have anticipated many interesting phenomena to occur in the presence of rotation, such as chiral vortical effect \cite{Kharzeev:2010gr,Kharzeev:2007tn}, chiral vortical wave \cite{Jiang:2015cva}, splitting of masses under rotation \cite{WeiMingHua:2020eee}, magnetic chiral density wave \cite{Ghalati:2023npr}, vortical effects in AdS space\cite{Ambrus:2021eod} etc. These are anomalous processes which depend on the experimental signatures can be observed in heavy-ion collision experiments. Apart from this the study of QCD phase diagram \cite{Becattini:2021lfq,Sadooghi:2021upd} under rotation is yet another interesting topic of research. Under the influence of large angular velocities, the QCD medium offers the scope to investigate anomalous properties in greater detail as they will be properly exhibited, thus allowing an easier extraction of the experimental signatures. Particularly in Ref. \cite{Chernodub:2020qah}, the effect of rotation has been studied on the confining and deconfining phases of QGP, where it is found that at particular distance from the rotation axis, the deconfining transition is observed. In addition to the usual confinement and deconfinement phases, there exists a mixed inhomogeneous phase, which contains spatially separated confinement and deconfinement regions, thus two deconfining temperatures are predicted to be present. Studies have also been performed for lattice SU(3) gauge theory \cite{Braguta:2023yjn}, where the isothermal moment of inertia for a rigidly rotating QGP has been calculated. It was found that the moment of inertia takes negative values below the supervortical temperature \cite{Braguta:2023yjn,Braguta:2023tqz}, which is 1.5 times the critical temperature, indicating a thermodynamic instability of rigid rotation beyond a certain temperature. In the more recent work of Ref. \cite{Braguta:2023tqz}, the origin of the negative values of the moment of inertia has been attributed to a novel effect known as the negative gluonic Barnett effect occuring due to the spin-vortical coupling for gluons, wherein the spin polarization of gluons exceeds the total angular momentum of the system, thus compelling the orbital angular momentum to take negative values in the supervortical range of temperatures. Studies concerning fermions, have been performed for Nambu-Jona-Lasinio model in Ref. \cite{Chernodub:2017ref}, by taking into account chiral MIT boundary conditions parameterized by a chiral angle $\Theta$, which basically confine the fermions in the cylinder. An interesting observation of Ref. \cite{Chernodub:2017ref} is that the chiral restoration temperature $T_c$ decreases quadratically with the angular velocity, while the position and slope of the critical curve depends on $\Theta$.\\
One of the topics concerned with rotating QCD matter is the study of transport coefficients \cite{Hu:2021lnx}, which enter as inputs for performing hydrodynamical simulations. There exist various formalisms to compute the transport coefficients viz. Kinetic theory \cite{Bhadury:2020cop,Bhadury:2020puc} studied through the Relativistic Boltzmann Equation (RBE), based on kinetic theory and Kubo formalism \cite{Hongo:2021ona,Harutyunyan:2018wdk,Harutyunyan:2021rmb,Czajka:2018bod,Becattini:2019dxo,Kovtun:2018dvd,Shukla:2019shf}, based on the calculation of correlation functions and spectral functions from quantum field theory. In this work we have calculated the bulk viscosity from correlation functions of fields \cite{Czajka:2018bod} which is a first principle approach. The Nonequilibrium Statistical Operator formalism (NESO) is a framework which allows to derive and compute transport coefficients by the help of Kubo formulas \cite{Harutyunyan:2018wdk,Harutyunyan:2021rmb,Becattini:2019dxo,Kovtun:2018dvd,Shukla:2019shf}. A recent study by Hu \cite{Hu:2021lnx} provides the Kubo formulas, derived from NESO formalism to calculate the transport coefficients of a rotating, hot and dense relativistic fluid. In order to approach this problem one has to calculate the correlation functions of fields which in turn allow us to compute the spectral function concerned with the corresponding transport coefficient. A similar study has also been perfomed by Hongo {\it et.al} in Ref. \cite{Hongo:2021ona} for QED and QCD in a torsionful background. By analysing the situation of the fermion propagator in magnetic field \cite{Chyi:1999fc}, one can guess that the fermion propagator in the presence of rotation will not remain translationally invariant anymore \cite{Chyi:1999fc,Ayala:2021osy} due to the fixing of the rotation axis along the $z$ direction. A situation reminiscent to this would be that of a background magnetic field wherein the external gauge field is coupled to other fields (fermionic and complex scalar fields). In reference \cite{Hongo:2021ona}, QCD has been studied in a torsionful background in which the fundamental variables describing the background geometry are the vierbein and the spin connection (or contorsion \cite{Hongo:2021ona}) which are independent background fields. With the introduction of torsion, the spin connection is promoted to an independent background field, which is equivalent to putting the system in a background with torsion. A similar study has been performed by Gallegos {et.al} \cite{Gallegos:2022jow}, where the background torsion tensor is shown to couple to the spin current. Thus the coupling occurs between the quarks and the background torsion through the spin connection. Subsequently, Euler-Lagrange equations are computed \cite{Hongo:2021ona} by varying the action with respect to the vierbein and spin connection. Contrary to the fermionic fields, for spin 1 particles such as photon and gluon, the background torsion does not couple to the gauge field due to the SU(3)$_\text{c}$ gauge invariance of the action \cite{Hongo:2021ona,Hehl:1976kj}. Background fields such as magnetic field and rotation break the translational symmetry of the system and this has been shown in Ref. \cite{Ayala:2021osy}, where the fermion propagator is derived by the help of Fock-Schwinger method. But the recovery of translational invariance \cite{Ayala:2021osy} at large angular velocities allows us to employ the propagator for a perturbative study of various quantities.\\ 
In the present work we have studied the bulk viscosity of spin 1/2 fermions in a rotating medium using the Kubo formalism which is based on the techniques of quantum field theory at finite temperature, at very large angular velocities. Bulk viscosity is important for various reasons, with the two most important ones for a hydrodynamic system being that, it measures the deviation of the system from conformality and the other one being that it enters as input into hydrodynamical simulations along with shear viscosity. In cosmology, the contribution coming from bulk viscosity has been suggested in Ref. \cite{Cheng:1991uu} to contribute to cosmological inflation and density fluctuations. A finite bulk viscosity is helpful in resolving the entropy generation problem\cite{Cheng:1991uu,Tawfik:2009mk}. The finiteness of the quantity is decisive in studying the evolution of the early universe, which is described by a isotropic and homogeneous Robertson-Walker metric \cite{Tawfik:2010pm,Tawfik:2010mb,Tawfik:2010bm}. In nuclear collisions, bulk viscosity is useful in explaining the radial flow and azimuthal anisotropy in heavy-ion collisions \cite{STAR:2017ykf}. Bulk viscosity has a role to play to understand the bulk viscous effects on the properties of heavy quarkonia \cite{Thakur:2020ifi}, where the effect is responsible for distorting the distribution functions of quarks and gluons respectively. It is an important quantity for the BES (Beam Energy Scan) program \cite{Bzdak:2019pkr}, where bulk viscous effects are expected as the system approaches a critical point \cite{Kharzeev:2007wb,Karsch:2007jc,Moore:2008ws}.  \\
This work is organized as follows. In Section \ref{F-Rot} we briefly review the behaviour of fermions in a rotating environment. Then we calculate the spectral function and the bulk viscosity of spin 1/2 fermions in Section \ref{spec} by using the concepts discussed in Section \ref{F-Rot}. We then discuss the results of our work in Section \ref{Res}. In the end we have provided the calculational details in Appendix \ref{App-A}.

\section{Fermions in a rigidly rotating environment}
\label{F-Rot}

In this section we discuss in short, the topic of fermions in a rotating environment. To study this system we will work in curved space where the metric tensor resembles a curved space time that is useful for describing the geometry of the region formed after the non-central collision which is rotating with an angular velocity $\Omega$ around the $z$-axis. The metric tensor $g^{\mu\nu}$ for the system under study is given by 
\bea
g_{\mu\nu} = \begin{pmatrix}
	1 - (x^2 + y^2)\Omega^2 & y\Omega & -x\Omega & 0\\
	y\Omega  &  -1 & 0 & 0  \\
	-x\Omega & 0 & -1 & 0 \\
	0 & 0 & 0 & -1 
\end{pmatrix}.
\label{F-Rot-1}
\eea 
The Dirac equation describing a fermion in cylindrical coordinates\cite{Ayala:2021osy,Fang:2021mou} is given by 
\bea
\big(i\wtg^\mu\widetilde{D}_\mu - m \big)\psi = 0
\label{F-Rot-2}
\eea  
where $\wtg^\mu$ are the gamma matrices in curved space which are spacetime-dependent, $\widetilde{D}_\mu$ is the covariant derivative in curved space and $m$ is the mass of the fermion. Here $\widetilde{D}_\mu$ is given by 
$$\widetilde{D}_\mu = \partial_\mu + \Gamma_\mu$$
where $\Gamma_\mu = \frac{1}{8}\omega_{\mu a b}\big[\gamma^a, \gamma^b\big]$ is the affine connection and $\omega_{\mu a b}$ is the spin connection. To calculate $\omega_{\mu a  b}$ one has to use the definition of vierbein (also called tetrad) and metric tensor $g_{\mu\nu}$ which are given by 
\bea
g_{\mu\nu} = \eta_{ab}e^a_{~\mu}e^b_{~\nu},~e_a^{~\mu} = \delta_a^{~\mu} - \delta_a^{~0}\delta_i^{~\mu}v_i,~~~e^a_{~\mu} = \delta^a_{~\mu} + \delta^a_{~i}\delta^0_{~\mu}v_i,~(a,\mu = 0, 1,2, 3,~i = 1,2,3),
\label{F-Rot-3}
\eea  
which in turn can be used to calculate $\omega_{\mu a b}$ and $\Gamma^\beta_{~\mu\nu}$ as
\bea
\omega_{\mu a b} = g_{\alpha\beta}e_a^{~\alpha}\big[\partial_\mu e_b^{~\beta} + \Gamma^\beta_{~\mu\nu}e_b^{~\nu}  \big],~~\text{where~~}\Gamma^\beta_{~\mu\nu} = \frac{g^{\beta\alpha}}{2}\big[ \partial_\nu g_{\alpha\mu} + \partial_\mu g_{\alpha\nu}- \partial_\alpha g_{\mu\nu}  \big].
\label{F-Rot-4}
\eea 
Here $\eta_{ab} = \text{diag}(1,-1,-1,-1)$ is the metric tensor in Minkowski space. In curved spacetime the energy-momentum tensor\cite{Fang:2021mou,Buzzegoli:2017cqy,Buzzegoli:2018wpy,Palermo:2021hlf} is given by 
\bea
T^{\mu\nu} =  \frac{i}{4}\big(\overline{\psi}\wtg^\mu \wtD^\nu\psi + \overline{\psi}\wtg^\nu \wtD^\mu\psi  \big) + \text{H.C}
\label{F-Rot-5}
\eea 
where $\psi$ and $\overline{\psi}$ are the Dirac field and its conjugate, H.C denotes the Hermitian conjugate. In a uniformly rotating frame, $\widetilde{D}^\mu$ is given by 
\bea
\widetilde{D}^\mu = \big( \partial_t - i\frac{\Omega\Sigma_3}{2}, -\partial_x, -\partial_y, -\partial_z \big)
\label{F-Rot-6}
\eea 
and the spacetime dependent gamma matrices in tetrad space are given by 
\bea
&&\wtg^0 = \gamma^0,~~ \wtg^1 = \gamma^1 + y\Omega\gamma^0,~~\wtg^2 = \gamma^2 - x\Omega\gamma^0,~~\wtg^3 = \gamma^3\nn \\
&&\wtg_0 = \gamma^0 - x\Omega\gamma^2 + y\Omega\gamma^1,~~\wtg_1 = -\gamma^1, \wtg_2 = -\gamma^2,~~ \wtg_3 = -\gamma^3
\label{F-Rot-6A}
\eea 
where $\Sigma_3 = \frac{i}{2}\big[\gamma^1, \gamma^2\big]$ and $\gamma^\mu$, $\mu = 0,1,2,3$ are the gamma matrices in Minkowski space. In heavy-ion collisions the direction of rotation is perpendicular to the reaction plane which is taken to be the $x$-$y$ plane for our work. Following the above mentioned calculations the free Lagrangian \cite{Wei:2021dib} with finite chemical potential $\mu$ for a medium rotating with constant angular velocity $\Omega$ is given by 
\bea
\mathcal{L} = \overline{\psi}\big[i\gamma^\mu\partial_\mu + \gamma^0\big( \Omega J_z + \mu \big)- m\big]\psi,
\label{F-Rot-7}
\eea 
where $J_z$ is the third component of the total angular momentum $\vec{J} = \vec{x}\times \vec{p} + \vec{S}$ where $\vec{S} = \frac{1}{2}\begin{pmatrix}
	\vec{\sigma} & 0\\
	0 & \vec{\sigma}
\end{pmatrix}$, $\vec{\sigma}$ are the Pauli matrices and $m$ is the mass of the fermion. The fermion propagator \cite{Ayala:2021osy} in momentum space derived from this Lagrangian by the help of Fock-Schwinger method \cite{Iablokov:2020upc} is given by 
\bea
S(p) &=& \frac{\big( p_0 + \frac{\Omega}{2} - p_z + ip_\perp   \big)\big(\gamma_0 + \gamma_3\big) + m\big(1 + \gamma_5\big)}{\big(p_0 + \frac{\Omega}{2}\big)^2 - \vec{p}^2 - m^2 + i\epsilon}\mathcal{O}^+ + \frac{\big( p_0 - \frac{\Omega}{2} + p_z - ip_\perp   \big)\big(\gamma_0 - \gamma_3\big) + m\big(1 + \gamma_5\big)}{\big(p_0 - \frac{\Omega}{2}\big)^2 - \vec{p}^2 - m^2 + i\epsilon}\mathcal{O}^-\nn\\
\label{F-Rot-8}
\eea 
where  $p_0$ is the temporal component, $p_z$ is the $z$ component which is parallel to the axis of rotation and $p_\perp = \sqrt{p_x^2 + p_y^2}$ is the transverse component of the 4-momentum and $\mathcal{O}^{\pm} \equiv \frac{1}{2}\big[1 \pm i\gamma^1\gamma^2\big]$. We have employed Eq. (\ref{F-Rot-8}) to perform calculations at finite temperature by the help of Imaginary Time Formalism(ITF). 
\section{Spectral function and bulk viscosity of fermionic system in a rotating medium}
\label{spec}

The bulk viscous pressure measures the deviation of the pressure from its equilibrium value, which causes the fluid to expand or compress. From first order dissipative hydrodynamics the bulk viscosity $\zeta$ quantifies the dissipation when expansion or compression occurs in the fluid. The calculation of $\zeta$ via correlation functions is given by the Kubo formula  \cite{Hu:2021lnx,Harutyunyan:2018wdk,Czajka:2018bod}, which is based on the calculation of spectral functions of energy momentum tensor $T^{\mu\nu}(x)$ and Noether current $J^\mu(x)$ relevant to each transport coefficient. The Kubo formulas for a relativistic fluid are derived from the Nonequilibrium Statistical Operator(NESO) formalism \cite{Harutyunyan:2018wdk,Harutyunyan:2021rmb} developed by Zubarev, where the nonequilibrium averages of different dissipative quantities such as shear-stress tensor $\pi^{\mu\nu}$, energy diffusion flux $q^\mu$ and charge diffusion flux $j^\mu(x)$ are calculated in a nonequilibrium scenario. The quantities $\pi^{\mu\nu}, q^\mu$ and $j^\mu$ carry $T^{\mu\nu}(x)$ and charge current $J^\mu(x)$ along with rank-4 and rank-2 projectors, that are projected on to the spectral functions. Here, $T^{\mu\nu}(x)$ and $J^{\mu}(x)$ can be calculated from the respective quantum field theory under study. We will be restricting ourselves to the study of spin 1/2 fermions of a $U(1)$ gauge theory. The estimation of $\zeta$ through correlation functions at one-loop is calculated by the Kubo formula,
\bea
\zeta = - \displaystyle{\lim_{q_0\to 0}}\frac{\rho_\zeta(q)}{q_0}
\label{spec-1}
\eea     
where $\rho_\zeta(q)$ is the spectral function for bulk viscosity calculated from two-point correlation function of energy-momentum tensor $T^{\mu\nu}$, $q$  is the 4-momentum and $q_0$ is the temporal component of the 4-momentum of the external gauge Boson, which is the photon in our case. The expression of $\rho_\zeta(q)$ is given by 
\bea
\rho_\zeta(q) = \text{Im}\Pi(q)
\label{spec-1A}
\eea  
where $\Pi_\zeta(q)$ is given by 
\bea
\Pi_\zeta(q) &=& i\int d^4r~ e^{iq\cdot r} \big\langle \mathcal{P}^*(r)\mathcal{P}^*(0) \big\rangle_R,
\label{spec-1B}
\eea 
where the notation $\big\langle\mathcal{O}_1(x)\mathcal{O}_2(y)\big\rangle_R$ denotes the retarded thermal average of operators (say $\mathcal{O}_1(x)$ and $\mathcal{O}_2(y)$) at spacetime points $x$ and $y$. Here $\Pi_\zeta(q)$ is calculated in cylindrical coordinates where $r = (t, \rho, \theta, z)$ and $\mathcal{P}^*(x)$, which is the total pressure of the nonequilibrium system  \cite{Teaney:2009qa} is given by 
\bea
\mathcal{P}^*(x) = -\frac{T^i_{~i}}{3} - c_s^2T^{00},
\label{spec-2}
\eea 
where $T^i_{~i}$ and $T^{00}$ are various components of $T^{\mu\nu}$ carrying spatial index $(i)$ and temporal index $(0)$ and $c_s^2$ is the speed of sound, which for our case is $\frac{1}{3}$ as we have considered a system of massless and non-interacting fermions. To further calculate Eq. (\ref{spec-1A}), one has to perform the Wick's contraction of $\psi(0)$ and $\overline{\psi}(r)$ i.e 
\bea
\wick{\c\psi(0)  \c\ovpsi(r)} = S(0,r) = \int\frac{d^4p}{(2\pi)^4}e^{-ip\cdot x}\big(-iS(p)\big),
\label{spec-6}
\eea    
where $S(0,r)$ is the fermion propagator in coordinate space and $S(p)$ is the fermion propagator in momentum space. The two-point function of energy-momentum tensors calculated from Wick's contraction given by Eq. (\ref{spec-6}) reads as follows

\bea
\langle T^{\mu\nu}(x)T^{\alpha\beta}(y)\rangle 
&=& -\frac{1}{16}\Big\langle\left[\ovpsi\wtg^{\{\mu} (\wtD^{\nu\}}\psi)  - (\wtD^{\{\mu} \ovpsi)\wtg^{\nu\}}\psi \right]_x \left[\ovpsi\wtg^{\{\alpha} (\wtD^{\beta\}}\psi) - (\wtD^{\{\alpha} \ovpsi)\wtg^{\beta\}}\psi \right]_y\Big\rangle \, \nn \\
&=& -\frac{1}{16}\Big[\big[ \wick{ \c2\ovpsi\wtg^{\{\mu} (\wtD^{\nu\}}\c1\psi) \big]_x\big[  \c1\ovpsi\wtg^{\{\alpha} (\wtD^{\beta\}}\c2\psi)  \big]_y} -  \wick{\big[\c2\ovpsi\wtg^{\{\mu} (\wtD^{\nu\}}\c1\psi)\big]_x\big[(\wtD^{\{\alpha} \c1\ovpsi)\wtg^{\beta\}}\c2\psi\Big]_y }    \nn \\ 
&& - \wick{\big[(\wtD^{\{\mu} \c2\ovpsi)\wtg^{\nu\}}\c1\psi  \big]_x\big[ \c1\ovpsi\wtg^{\{\alpha} (\wtD^{\beta\}}\c2\psi)   \big]_y} + \wick{\big[  (\wtD^{\{\mu} \c2\ovpsi)\wtg^{\nu\}}\c1\psi  \big]_x\big[  (\wtD^{\{\alpha} \c1\ovpsi)\wtg^{\beta\}}\c2\psi  \big]_y}  \Big]\,,
\label{spec-2B}
\eea  

where $A^{\{\alpha\beta\}} = A^{\alpha\beta} + A^{\beta\alpha}$. In Eq. (\ref{spec-2B}) we have considered the general spacetime points $x$ and $y$, which can be specified for cylindrical coordinates at the points $r$ and $r'$. On substituting Eq. (\ref{spec-2B}) in Eq. (\ref{spec-8}), the general expression for the two-point function of energy-momentum tensors in cylindrical coordinates is given by 

\bea
\big\langle T^{\mu\nu}(0)T^{\alpha\beta}(r) \big\rangle 
&=& -\frac{1}{16}\Big[\Tr\big\{ \wtg^{\{\mu}\wtD^{\nu\}} S(0,r)\wtg^{\{\alpha} \wtD^{\beta\}} S(r,0)  \big\} - \Tr\big\{\wtg^{\{\mu} \wtD^{\nu\}}\wtD^{\{\alpha} S(0,r)\wtg^{\beta\}} S(r,0)  \big\} \nn \\
&& - \Tr\big\{ \wtg^{\{\mu} S(0,r)\wtg^{\{\alpha}\wtD^{\beta\}}\wtD^{\nu\}} S(r,0)   \big\}  + \Tr\big\{ \wtg^{\{\mu}\wtD^{\{\alpha} S(0,r)\wtg^{\beta\}} \wtD^{\nu\}} S(r,0)\big\}   \Big].
\label{spec-2C} 
\eea 

It should be noted that in a rotating environment, the translational invariance of the coordinate space propagator is lost \cite{Ayala:2021osy}, due to the preferred direction of the rotation axis along the $z$ direction. The breaking of translational invariance of the propagator along transverse directions the  has also been seen in a background magnetic field \cite{Kuznetsov:2004tb,Kuznetsov:2013sea,Chyi:1999fc}, where a phase factor shift appears in the expression of the propagator. However for a rotating medium, the translational invariance is recovered by assuming that the fermion is completely dragged by the vortical motion or large $\Omega$ which is of the order $10^{-1}$ GeV or $10^{22}$ sec$^{-1}$. In this situation the phase factor appearing in the coordinate space propagator \cite{Ayala:2021osy} disappears completely, thus making the expression of the propagator depend only on the relative coordinates i.e
$$S(r,r') \xrightarrow{\text{Large~} \Omega} S(r-r'),$$
which leads to a recovery of translational invariance. This situation is valid for early stages of heavy-ion collisions when the $\Omega$ produced is very large. In Eq. (\ref{spec-6}) we have represented the coordinate space propagator as a Fourier transformation of the momentum space propagator \cite{Ayala:2021osy} which is explicitly given by Eq. (\ref{F-Rot-8}). This can be used to write the propagators appearing in Eq. (\ref{spec-2C}) in terms of their Fourier transformed versions, thus applicable to the calculation of spectral functions. On using Eqs.(\ref{spec-2C}) in Eq. (\ref{spec-1B}), expression of $\Pi_\zeta(q)$ is given by
\begin{figure}[h]
\centering
\includegraphics[scale = 0.16]{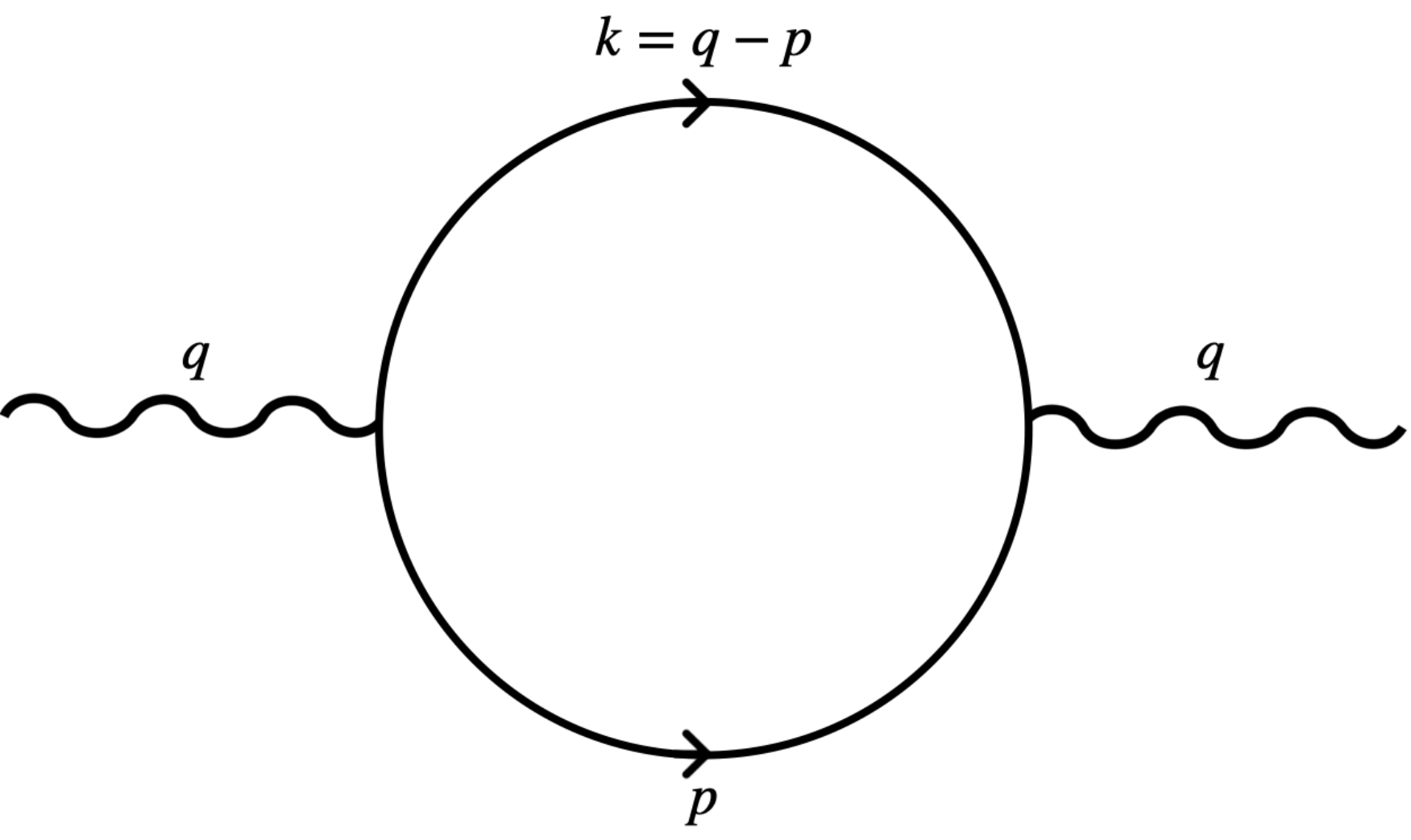}
\caption{One loop diagram representing the photon polarization tensor with $q$ being the momentum of the photon and $q-p$ and $p$ being the momentum of the antifermion and fermions respectively.}
\label{fig:oneloop}
\end{figure}
\bea
&&\Pi_\zeta(q) = i\int d^4r~e^{iq\cdot r}\big\langle \mathcal{P}^*(r)\mathcal{P}^*(0) \big\rangle_R  \nn \\
&& =-\frac{i}{4}\int d^4r\int\int\frac{d^4k}{(2\pi)^4}\frac{d^4p}{(2\pi)^4}e^{i(q-p-k)\cdot r}\Big[ (p_z + k_z)^2\Tr\big\{\wtg^3 S(p)\wtg^3S(k)\big\} \Big]\nn \\ 
&&+ i\frac{c_s^2}{12}\int d^4r\int\int\frac{d^4k}{(2\pi)^4}\frac{d^4p}{(2\pi)^4}e^{i(q-p-k)\cdot r} \Big[  \Tr\big\{ \wtg^3p_z S(p)\wtg^0\Big(k_0 + \frac{\Omega\Sigma_3}{2}\Big) S(k) \big\}  - \Tr\big\{\wtg^3p_z\Big(p_0 + \frac{\Omega\Sigma_3}{2}\Big)S(p)\wtg^0S(k)\big\} \nn\\ &&  + \Tr\big\{\wtg_3S(p)\wtg^0k_z\Big(k_0 + \frac{\Omega\Sigma_3}{2}\Big)S(k)\big\} - \Tr\big\{\wtg_3\Big(p_0 + \frac{\Omega\Sigma_3}{2}\Big)S(p)k_z\wtg^0S(k)\big\}   
+ \Tr\big\{ \wtg^0 \Big(p_0 + \frac{\Omega\Sigma_3}{2}S(p)\wtg^3S(k)\Big)  \big\} \nn \\ &&+ \Tr\big\{ \wtg^0\Big(p_0 + \frac{\Omega\Sigma_3}{2}\Big)S(p)\wtg^3S(k)  \big\}   - \Tr\big\{\wtg^0S(p)k_z\wtg^3\Big(k_0 + \frac{\Omega\Sigma_3}{2}\Big)S(k) \big\} - \Tr\big\{\wtg^0 p_zS(p)\wtg^3\Big(k_0 + \frac{\Omega\Sigma_3}{2}\Big)S(k)\big\}      \Big] \nn \\
&&+ \frac{i}{2}\Big(\frac{c_s^2}{3}\Big)^2\int d^4r\int\int\frac{d^4k}{(2\pi)^4}\frac{d^4p}{(2\pi)^4}e^{i(q-p-k)\cdot r}\Big\{ \Tr\big\{\wtg^0\Big(p_0 + \frac{\Omega\Sigma_3}{2})S(p)\wtg^0\Big(p_0 + \frac{\Omega\Sigma_3}{2}\Big)S(k)   \big\} \nn \\ && + \frac{1}{2}\Tr\big\{\wtg^0\Big(p_0 + \frac{\Omega\Sigma_3}{2}\Big)^2S(p)\wtg^0S(k)\big\}    + \frac{1}{2}\Tr\big\{\wtg^0S(p)\wtg^0\Big(k_0 + \frac{\Omega\Sigma_3}{2}\Big)^2S(k)\big\} \Big\}
\label{spec-8A}
\eea 

On employing the Wick's contraction given in Eq. (\ref{spec-6}) we proceed to evaluate Eq. (\ref{spec-8A}). The integration over $d^4r$ and $d^4k$ gives the delta function $\delta^{(4)}(q-p-k)$ on account of momentum conservation at the vertex of the photon polarization tensor, which fixes the relation between $q, p$ and $k$ as $k_\perp = q_\perp - p_\perp, k_z = q_z - p_z$ and $k_0 = q_0 - p_0$ in accordance with the one-loop diagram representing photon polarization in Fig. \ref{fig:oneloop}. Following Eq. (\ref{spec-1B}), $\rho_\zeta(q)$ is given by 
\bea
&&\rho_\zeta(q) = \text{Im}\Pi_\zeta(q)  \nn \\
&& =-\text{Im}\frac{i}{4}\int\frac{d^4p}{(2\pi)^4}  \Big[ q_z^2 \Tr\big\{\wtg^3 S(p)\wtg^3S(q-p)\big\}    \Big]\nn \\
&& + \text{Im}i\frac{c_s^2}{12}\int\frac{d^4p}{(2\pi)^4} \Big[  \Tr\big\{ \wtg^3p_z S(p)\wtg^0\Big(q_0 - p_0 + \frac{\Omega\Sigma_3}{2}\Big) S(q-p) \big\} - \Tr\big\{\wtg^3p_z\Big(p_0 + \frac{\Omega\Sigma_3}{2}\Big)S(p)\wtg^0S(q-p)\big\} \Big] \nn \\
&& + \Tr\big\{\wtg_3S(p)\wtg^0(q_z - p_z)\Big(q_0 - p_0 + \frac{\Omega\Sigma_3}{2}\Big)S(q-p)\big\} - \Tr\big\{\wtg_3\Big(p_0 + \frac{\Omega\Sigma_3}{2}\Big)S(p)(q_z - p_z)\wtg^0S(q-p)\big\}    \nn \\
&&+ \Tr\big\{ \wtg^0 \Big(p_0 + \frac{\Omega\Sigma_3}{2}\Big)S(p)\wtg^3S(q-p)\Big)  \big\} + \Tr\big\{ \wtg^0\Big(p_0 + \frac{\Omega\Sigma_3}{2}\Big)S(p)\wtg^3S(q-p)  \big\}  \nn \\
&& - \Tr\big\{\wtg^0S(p)(q_z - p_z)\wtg^3\Big(q_0 - p_0 + \frac{\Omega\Sigma_3}{2}\Big)S(q-p) \big\} - \Tr\big\{\wtg^0 p_zS(p)\wtg^3\Big(q_0 - p_0 + \frac{\Omega\Sigma_3}{2}\Big)S(q-p)\big\}      \Big] \nn \\
&&+ \text{Im}\frac{i}{2}\Big(\frac{c_s^2}{3}\Big)^2\int\frac{d^4p}{(2\pi)^4}\Big[ \Tr\big\{\wtg^0\Big(p_0 + \frac{\Omega\Sigma_3}{2})S(p)\wtg^0\Big(p_0 + \frac{\Omega\Sigma_3}{2}\Big)S(q-p)   \big\}\nn \\ && + \frac{1}{2}\Tr\big\{\wtg^0\Big(p_0 + \frac{\Omega\Sigma_3}{2}\Big)^2S(p)\wtg^0S(q-p)\big\} 
+ \frac{1}{2}\Tr\big\{\wtg^0S(p)\wtg^0\Big(q_0 - p_0 + \frac{\Omega\Sigma_3}{2}\Big)^2S(q-p)\big\} \Big]
\label{spec-8}
\eea 
In order to calculate Eq. (\ref{spec-8}) we will use ITF, where we will perform Matsubara frequency summation at finite chemical potential and angular velocity. The details on these computations has been provided in Appendix \ref{App-A}. On performing Matsubara frequency summation at finite $T, \mu$ and $\Omega$, we see that
in Eqs.(\ref{App-A-6})-(\ref{App-A-9}) we have four combinations of $q_0, E_k$ and $E_p$ appearing in the denominator as $q_0 \pm E_k \pm E_p$ and $q_0 \mp E_k \pm E_p$ which can be considered to be the limiting values of energy from the upper half plane. By setting the Bosonic Matsubara frequency (see Appendix \ref{App-A}) $\nu_N \to -i(q_0 + i0^+)$ via analytic continuation of discrete Matsubara frequencies to continuous energies, we can take the imaginary part as required in Eq. (\ref{spec-8}) by using the relation
\bea
\frac{1}{\Delta \pm i0^+} = \mathbb{P}\Big(\frac{1}{\Delta}\Big) \mp i\pi \delta(\Delta)
\label{spec-13}
\eea  
where $\mathbb{P}(x)$ is the principal part of the expression. On doing this the expression of $\rho_\zeta(q)$ reads as 
\bea
&&\rho_\zeta(q)\nn \\ &&= - \pi \int\frac{d^3p}{(2\pi)^3}\Big\{\frac{\mathcal{A}(p_\perp, p_z, p_0, q_\perp, q_z, q_0) + (c_s^2/3)\mathcal{C}(p_\perp, p_z, p_0, q_\perp, q_z, q_0) + (c_s^4/18)\mathcal{E}(p_\perp, p_z, p_0, q_\perp, q_z, q_0)}{4E_pE_{q-p}}\Big\} \nn \\ &&\Big[\Big\{n_F(E_{q-p} + \widetilde{\mu}) - n_F(E_p + \widetilde{\mu})\Big\}\delta(q_0 + E_{q-p} - E_p) \nn \\
&&+ \Big\{n_F(E_p - \widetilde{\mu}) - n_F(E_{q-p} - \widetilde{\mu})\Big\}\delta(q_0 + E_p - E_{q-p}) \nn \\ && +  \Big\{ n_F(E_{q-p} + \widetilde{\mu} ) + n_F(E_p - \widetilde{\mu}) -1  \Big\}\delta(q_0 - E_{q-p} - E_p) \nn \\
&&+ \Big\{ 1-n_F(E_{q-p} - \widetilde{\mu} ) - n_F(E_p + \widetilde{\mu})   \Big\}\delta(q_0 + E_{q-p} + E_p)  \Big]\nn \\
&&-\pi\int\frac{d^3p}{(2\pi)^3}\Big\{\frac{\mathcal{B}(p_\perp, p_z, p_0, q_\perp, q_z, q_0) + (c_s^2/3)\mathcal{D}(p_\perp, p_z, p_0, q_\perp, q_z, q_0) + (c_s^4/18)\mathcal{F}(p_\perp, p_z, p_0, q_\perp, q_z, q_0)}{4E_pE_{q-p}}\Big\} \nn \\ &&\Big[\Big\{n_F(E_{q-p} + \bar{\mu}) - n_F(E_p + \bar{\mu})\Big\}\delta(q_0 + E_{q-p} - E_p) \nn \\
&& + \Big\{n_F(E_p - \bar{\mu}) - n_F(E_{q-p} - \bar{\mu})\Big\}\delta(q_0 + E_p - E_{q-p}) \nn \\ && + \Big\{ n_F(E_{q-p} + \bar{\mu} ) + n_F(E_p - \bar{\mu}) -1  \Big\}\delta(q_0 - E_{q-p} - E_p) \nn \\
&& + \Big\{ n_F(E_{q-p} - \bar{\mu} ) + n_F(E_p + \bar{\mu}) -1  \Big\}\delta(q_0 + E_{q-p} + E_p)  \Big], \nn \\
\label{spec-81}
\eea 
where the functions coming from the trace of $\wtg^\mu$ matrices i.e $\mathcal{A}(p_\perp, p_z, p_0, q_\perp, q_z, q_0)$, $\mathcal{B}(p_\perp, p_z, p_0, q_\perp, q_z, q_0)$,\\ 
$\mathcal{C}(p_\perp, p_z, p_0, q_\perp, q_z, q_0)$, $\mathcal{D}(p_\perp, p_z, p_0, q_\perp, q_z, q_0)$, $\mathcal{E}(p_\perp, p_z, p_0, q_\perp, q_z, q_0)$ and $\mathcal{F}(p_\perp, p_z, p_0, q_\perp, q_z, q_0)$, are given by
\bea
&&\mathcal{A}(p_\perp, p_z, p_0, q_\perp, q_z, q_0) =   -4\big\{(q_z - p_z)^2 + (q_z - p_z) p_z + p_z^2\big\}\big\{(q_0 - p_0 + q_z - p_z)(4p_0 + 4p_z - 2\Omega) \nn \\   && \hspace{3.9cm} - 4(q_\perp - p_\perp) p_\perp- 2(p_0 + p_z )\Omega + \Omega^2 \big\} \\
&&\mathcal{B}(p_\perp, p_z, p_0, q_\perp, q_z, q_0) = -4\big\{(q_z - p_z)^2 + (q_z - p_z) p_z + p_z^2\big\}\big\{(q_0 - p_0 + q_z - p_z)(4p_0 - 4p_z + 2\Omega)\nn \\  && \hspace{3.9cm} - 4(q_\perp - p_\perp) p_\perp  + 2(p_0 - p_z)\Omega + \Omega^2 \big\} \\
&&\mathcal{C}(p_\perp, p_z, p_0, q_\perp, q_z, q_0) =  \big\{4(q_0 - p_0)(q_z ) - (q_z - p_z) \Omega - 4p_0 p_z\big\}\big\{( q_z - p_z + q_0 - p_0)(4p_0 + 4p_z - 2\Omega) \nn \\ &&\hspace{3.8cm}- 4(q_\perp - p_\perp) p_\perp - 2(p_0 + p_z)\Omega + \Omega^2  \big\} \\
&&\mathcal{D}(p_\perp, p_z, p_0, q_\perp, q_z, q_0) = \big\{4(q_0 - p_0)(3q_z - 3p_z ) - p_z\Omega + 4p_0 p_z\big\}\big\{( q_z- p_z -q_0 + p_0)(4p_0 - 4p_z + 2\Omega) \nn \\ &&\hspace{3.8cm}+ 4(q_\perp - p_\perp) p_\perp - 2(p_0 - p_z)\Omega - \Omega^2  \big\}\\
&&\mathcal{E}(p_\perp, p_z, p_0, q_\perp, q_z, q_0) = -\frac{1}{2}\big\{2(q_0 - p_0) p_0 - 3(q_0 - p_0)\Omega - 4p_0^2 \big\}\big\{(q_0 - p_0- q_z + p_z)(4p_0 - 4p_z + 2\Omega) \nn \\ 
&&\hspace{3.8cm}- 4(q_\perp - p_\perp) p_\perp + 2(p_0 - p_z)\Omega + \Omega^2   \big\}\\
&&\mathcal{F}(p_\perp, p_z, p_0, q_\perp, q_z, q_0) = -\frac{1}{2}\big\{2(q_0 - p_0) p_0 + 3(q_0 - p_0)\Omega - 4p_0^2 \big\}\big\{(q_0 - p_0+ q_z - p_z)(4p_0 + 4p_z - 2\Omega) \nn \\ 
&&\hspace{3.8cm}- 4(q_\perp - p_\perp) p_\perp - 2(p_0 + p_z)\Omega + \Omega^2   \big\},
\eea 
and $\overline{\mu} = \mu + \Omega / 2$ and $\widetilde{\mu} = \mu - \Omega / 2$. It should be noted that $\rho_\zeta(q)$ given by Eq. (\ref{spec-81}), has been calculated from an analytic continuation of Matsubara frequencies and the cut of an Euclidean correlator represents real scatterings of on-shell particles which are kinematically allowed. Therefore physically, the imaginary parts correspond to real scatterings of particles whose information is carried by the delta functions appearing in Eq. (\ref{spec-81}). The delta functions $\delta(q_0 \pm E_p \mp E_k)$ are known as Landau cuts \cite{Mallik:2016anp} and $\delta(q_0 \pm E_p \pm E_k)$ are known as unitary cuts \cite{Mallik:2016anp}. The Landau cuts appear only at finite temperature whereas the unitary cuts are already present due to vacuum contributions. The unitary cuts do not contribute to the calculation of transport coefficients \cite{Mallik:2016anp} and are thus ignored in our calculations. From Eq. (\ref{spec-1}) the bulk viscosity is given by
\bea
&&\zeta(T,\mu,\Omega) \nn \\ &&= \displaystyle{\lim_{ q_0 \to 0, \vec{q}=\vec{0}}}\text{Im}\Bigg[-\pi\int\frac{d^3p}{(2\pi)^3}\frac{1}{4E_{q-p}E_p}\Big\{ \mathcal{A}(p_\perp, p_z, p_0, q_\perp, q_z, q_0)+ \Big(\frac{c_s^2}{3}\Big)\mathcal{C}(p_\perp, p_z, p_0, q_\perp, q_z, q_0) \nn \\ &&+ \Big(\frac{c_s^4}{18}\Big)\mathcal{E}(p_\perp, p_z, p_0, q_\perp, q_z, q_0)\Big\}\displaystyle{\lim_{\Gamma\to 0}}\Bigg\{\frac{\big(n_F(E_{q-p} + \bar{\mu}) - n_F(E_p + \bar{\mu})\big)/q_0}{q_0 + E_{q-p} - E_p + i\Gamma} + \frac{\big(n_F(E_{q-p} - \bar{\mu}) - n_F(E_p - \bar{\mu})\big)/q_0}{q_0 + E_p - E_{q-p} + i\Gamma} \Bigg\}\Bigg]\nn \\
&&+\hspace{-0.2cm}\displaystyle{\lim_{ q_0 \to 0, \vec{q}=\vec{0}}}\text{Im}\Bigg[-\pi\int\frac{d^3p}{(2\pi)^3}\frac{1}{4E_{q-p}E_p}\Big\{ \mathcal{B}(p_\perp, p_z, p_0, q_\perp, q_z, q_0) + \Big(\frac{c_s^2}{3}\Big)\mathcal{D}(p_\perp, p_z, p_0, q_\perp, q_z, q_0)\nn \\ && + \Big(\frac{c_s^4}{18}\Big)\mathcal{F}(p_\perp, p_z, p_0, q_\perp, q_z, q_0)\Big\}\displaystyle{\lim_{\Gamma\to 0}}\Bigg\{\frac{\big(n_F(E_{q-p} + \widetilde{\mu}) - n_F(E_p + \widetilde{\mu})\big)/q_0}{q_0 + E_{q-p} - E_p + i\Gamma} + \frac{\big(n_F(E_{q-p} - \widetilde{\mu}) - n_F(E_p - \widetilde{\mu})\big)/q_0}{q_0 + E_p - E_{q-p} + i\Gamma} \Bigg\}\Bigg],\nn \\
\label{spec-15}
\eea 
where we have substituted the Breit-Wigner form of the delta function i.e $\delta(x) = -\pi^{-1}\displaystyle{\lim_{\Gamma\to 0}}\frac{1}{x + i\Gamma}$. The parameter $\Gamma$ introduced here is the thermal width of the medium arising from interactions in the medium at finite temperature, where the temperature of the system being very high acts effective enough to generate interactions (i.e a finite $\Gamma$) and is basically the inverse of relaxation time $\tau$ i.e $\Gamma = \tau^{-1}$ \cite{Mallik:2016anp}. It should be noted that in the present work $\Gamma$, which has been introduced through the Breit-Wigner form can be similarly understood by introducing half-width $i\Gamma/2$  in the momentum space propagators. This can be done by considering the expression of the inverse of the retarded momentum space propagator $G^{-1}_{\text{R}}(p)$ which includes the complex valued self-energy $\Sigma_\text{R}(p)$, given by 
$G_{\text{R}}^{-1}(p) = p^2 - m_0^2 - \text{Re}\Sigma_{\text{R}}(p) - i\text{Im}\Sigma_{\text{R}}(p)\sim (p_0 + i\gamma(p))^2 - E_p^2$, where $\gamma = -\frac{1}{2p_0}\text{Im}\Sigma_{\text{R}}(p) = \frac{\Gamma}{2} $. This shows that the information of $\Gamma$ is already present in the propagator of the theory. For our case, we have employed a particular representation of the $\delta(x)$ function to inoculate this information in our calculations. One can calculate $\Gamma$ by considering an interaction Lagrangian and employing the techniques of thermal field theory \cite{Laine:2016hma,Lang:2012tt}. We see that in the $\vec{q} =\vec{0}, q_0 \to 0$ limit, $\zeta$ takes an indeterminate $0/0$ form, which is unphysical. Therefore we apply the L'Hospital rule to treat this divergence. By turning off the interaction i.e $\Gamma \to 0$, which happens for a free theory, the bulk viscosity diverges. But at finite temperature, excitations cannot propagate indefinitely in the medium without collisions, thus acquiring a finite lifetime which is inversely proportional to $\Gamma$ of the medium. Therefore one has to keep $\Gamma$ finite for obtaining a non-divergent transport coefficient. From the definition of $\zeta$ given in Eq. (\ref{spec-1}), we take the long wavelength limit or the $\vec{q} = \vec{0}, q_0 \to 0$ limit of $\frac{\rho_\zeta(\vec{q},q_0)}{q_0}$. Thus we obtain $\zeta(T,\mu,\Omega)$ from $\rho_\zeta(q)$, which is given by 
\bea
&&\zeta(T,\mu, \Omega)  \nn \\ &&=\frac{1}{8T}\Bigg\{\int\frac{d^3p~p_z^2}{(2\pi)^3~4E_p^2\Gamma}\Big[\Big\{4\big(E_p + p_z)^2 - 4p_\perp^2 -  \Omega^2\Big\} \mathcal{N}(\pm\mu, \pm\Omega/2) +  \Big\{4\big(E_p - p_z)^2 - 4p_\perp^2 -  \Omega^2\Big\} \mathcal{N}(\pm\mu, \mp\Omega/2)\Big]\Bigg\}\nn \\ 
&& + \frac{c_s^2}{12T}\Bigg\{\int\frac{d^3p}{(2\pi)^3~4E_p^2\Gamma} \Big[\Big\{ -2p_z\Big(-2E_p^3 + E_p^2\big(\Omega - 8p_z\big) + 2E_p\big(\Omega^2 + p_\perp^2 + 2p_z\big(\Omega - p_z\big)\big)   \nn \\ && + 2\Omega\big(p_z^2 - p_\perp^2\big)   \Big) \Big\}  \mathcal{N}(\pm \mu,\pm \Omega/2) - \frac{\Omega p_z}{2}\Big\{8E_p^2 - 4p_\perp^2 + 8E_pp_z + 3p_z^2    \Big\}\mathcal{N}(\pm\mu,\mp\Omega/2) \Big]\Bigg\} \nn \\
&&+ \frac{c_s^4\Omega}{144T}\Bigg\{ \int\frac{d^3p}{(2\pi)^3~4E_p^2\Gamma}\Big[ \Big\{\Omega^3 - 4\Omega \Big(E_p^2 + p_\perp^2 - p_z^2 \Big) + E_p\Big(3\Omega^2 + 4p_z\big(\Omega + p_z\big)\Big)   \Big\}\mathcal{N}(\pm \mu,\pm \Omega/2) \nn \\
&& + \Big\{ \Omega^3 + 8\Omega\big( p_\perp^2 - 2E_p^2  \big) - 4\big(2\Omega + E_p\big)p_z^2   \Big\}\mathcal{N}(\pm \mu,\mp \Omega/2)\Big]  \Bigg\}
\label{spec-9}
\eea

where $\mathcal{N}(\pm\mu, \pm\Omega/2)$ and $\mathcal{N}(\pm\mu, \mp\Omega/2)$ are given by 

\bea
&&\mathcal{N}(\pm\mu, \pm\Omega/2) = \sum_{s = \pm 1}n_F(E_p + s\mu + s\Omega/2)\big(1 - n_F(E_p + s\mu + s\Omega/2)\big)
\nn \\
&&\mathcal{N}(\pm\mu, \mp\Omega/2) = \sum_{s = \pm 1}n_F(E_p + s\mu - s\Omega/2)\big(1 - n_F(E_p + s\mu - s\Omega/2)\big),
\label{spec-10} \nn \\
\eea

where $n_F(x)$ is the Fermi-Dirac distribution function, $E_p = \sqrt{p_z^2 + p_\perp^2 + m^2}$ is the energy.
The final expression of $\zeta$ for a rotating, hot and dense fermionic system as a function of $T,\mu$ and $\Omega$, obtained from Kubo formalism is given by Eq. (\ref{spec-9}). It is clear from Eq. (\ref{spec-10}) that $\zeta$ has a depnedence on $\Omega$ in the distribution function and as well as in the numerator of the expression of $\zeta$. It does influence the effect of $\mu$, but at the same time this does not exactly behave like $\mu$, which is clear from the functional dependence of $\zeta$ in Eq. (\ref{spec-10}). Therefore it acts as an effective chemical potential in the medium. A hint of rotation and density acting as effective versions \cite{Chen:2015hfc} of each other is seen in non-relativistic theories when the hamiltonian $H$ is shifted to $H - \vec{L}.\vec{\Omega}$, where the latter term can be considered as an effective chemical potential. For our case which is a relativistic set-up, it so happens that $\Omega$ automatically enters in to the distribution function without any priori assumption, through Matsubara frequency summation. In a relativistic setting we see the effect of quantum field theory in curved spacetime in play, bringing in the effect of rotation into the picture, which is encoded in the metric of the cylindrical coordinates in which the Dirac equation \cite{Ayala:2021osy} has been solved.


\section{Results and discussions}
\label{Res}

In this section we have studied the behaviour of bulk viscosity as a function of $\Omega$ between 0.1 to 1 GeV, estimated from a one-loop diagram of the photon polarization in the limit of zero momentum and vanishing frequency of the spectral density, involving the two-point function of energy-momentum tensor. We should note that there are three energy scales associated with the system i.e $T, \mu$ and $\Omega$, all of the dimension GeV, which together dictate the behaviour of $\zeta$. Here we have mainly focussed on the purely deconfined phase that falls in the temperature range of $T$ = 0.2 to 0.4 GeV.

A complete pragmatic study of $\zeta$ from field theory will require the resummation of an infinite number of Feynman diagrams known as ladder diagrams \cite{Jeon:1994if,Jeon:1995zm,Weiner:2022kgx,Aarts:2002tn,Aarts:2004sd,Romatschke:2023ztk,Romatschke:2021imm,Romatschke:2019ybu,Romatschke:2019wxc,Romatschke:2019rjk,Romatschke:2019wxc,Romatschke:2019qbx,Lang:2012tt}, going beyond the evaluation of one-loop diagram. The resummation helps in the regulation of pinching poles that appear in the $\Gamma \to 0$ limit when the coupling $g$ is very small i.e $g << 1$. For a one-loop diagram studied in this work, the transport coefficient scales as $1/\Gamma$. The scaling for a two-loop diagram \cite{Aarts:2002tn,Aarts:2004sd,Lang:2012tt} is of the form $g^2 / \Gamma^2$ and subsequently for a $n$-loop diagram the scaling is of the form $g^{2n} / \Gamma^{n+1}$. The behaviour of the scaling shows that for every additional rung in the ladder diagram there is a dimensionful scaling factor of $g^2 / \Gamma$. Since $\Gamma \sim g^{2}$, for every $n$ the ladder diagram is of the order $\mathcal{O}(g^{-2})$, thus requiring a resummation. But nevertheless, a one-loop Kubo estimation of transport coefficients serves as a starting point to compute and analyse the behaviour of transport coefficients.

\begin{figure}[h]
	\includegraphics[scale = 0.22]{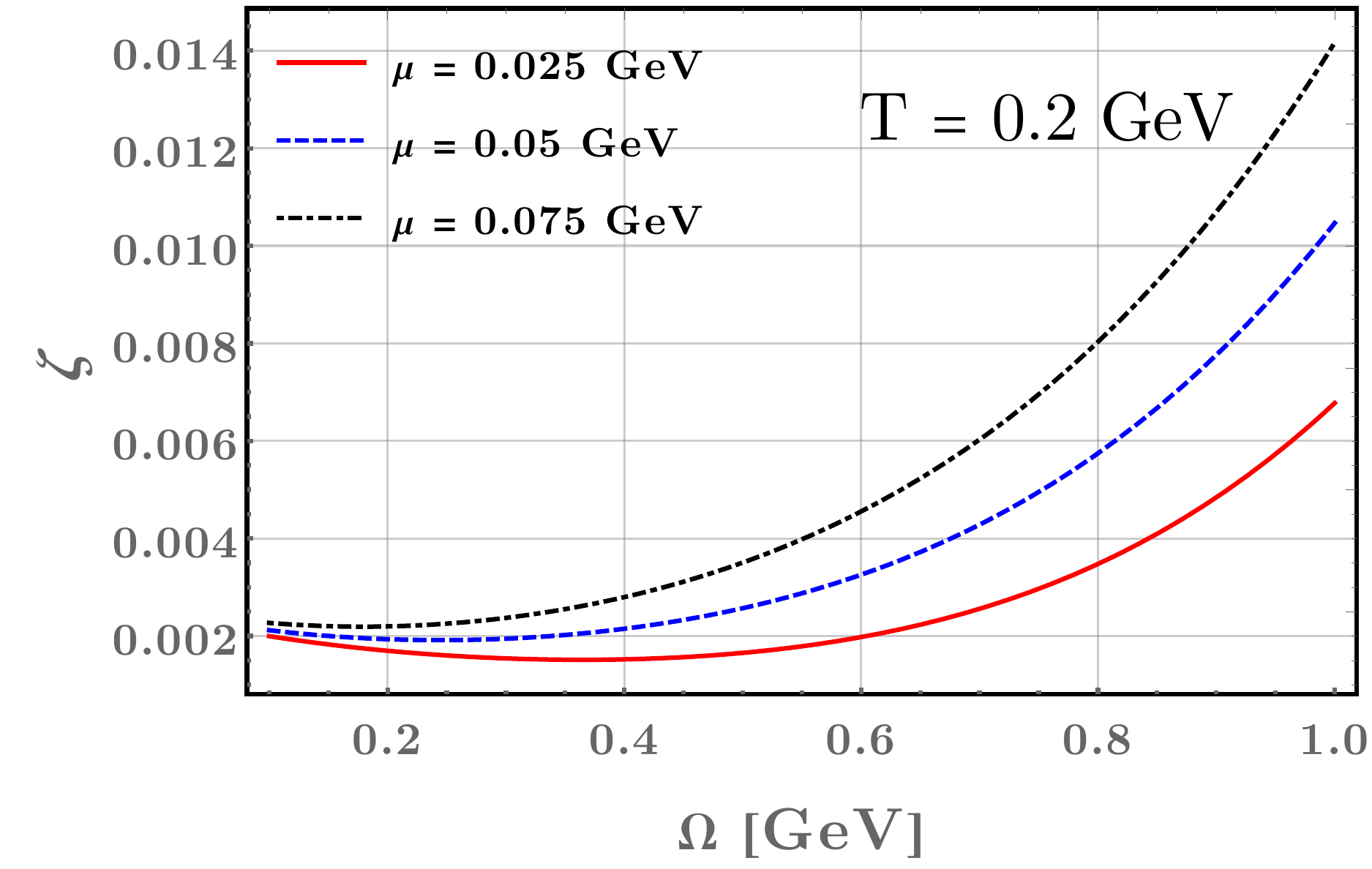}
	\caption{Plot of $\zeta$ vs $\Omega$ at $T =$ 0.2 GeV for $\mu = 0.025, 0.05$ and $0.075$ GeV.}
	\label{Zeta_Omg-1}
\end{figure}

\begin{figure}[h]
	\includegraphics[scale = 0.22]{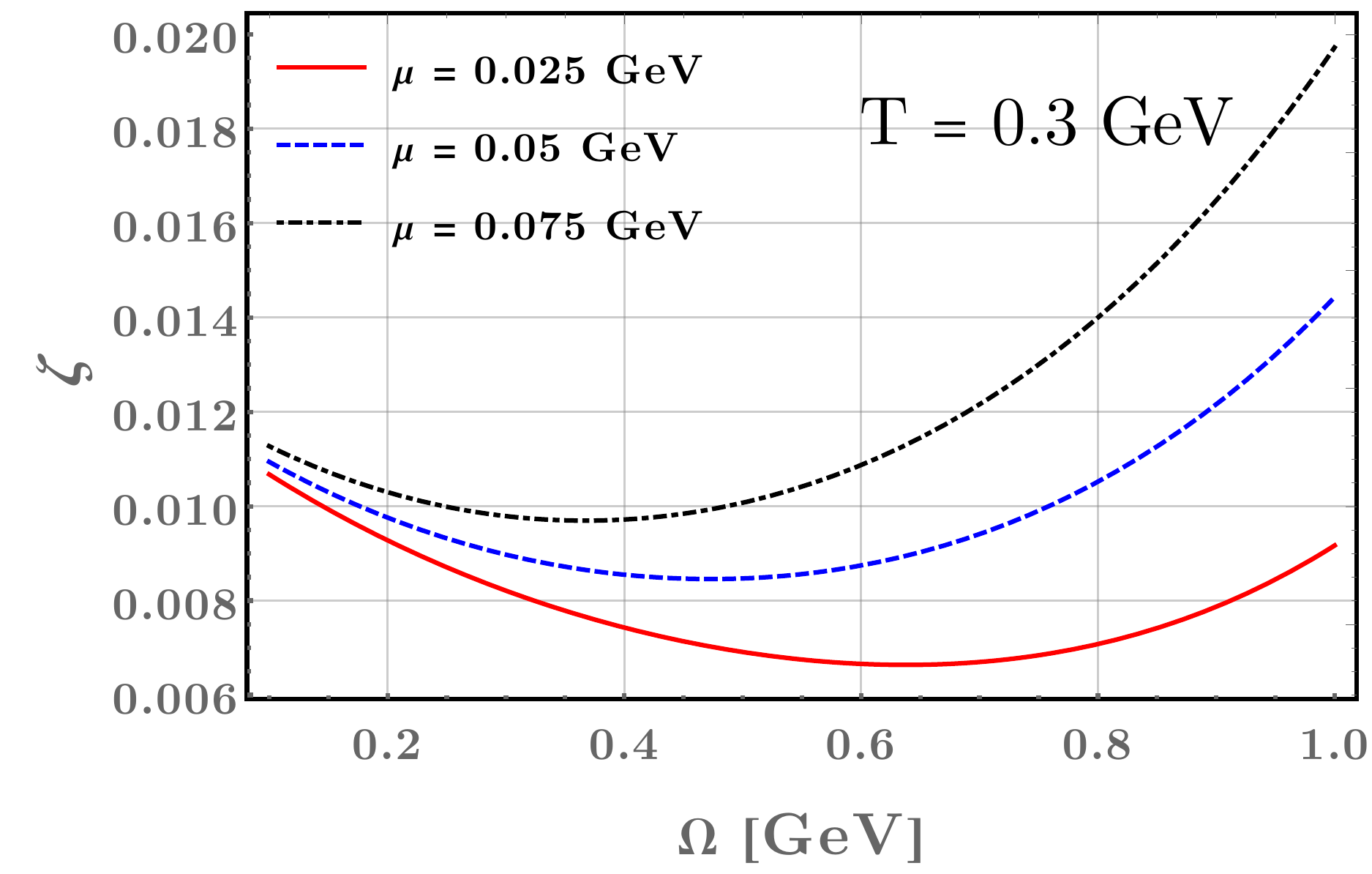}
	\caption{Same as fig. \ref{Zeta_Omg-1}, but for $T =$ 0.3 GeV}
	\label{Zeta_Omg-2}
\end{figure}

\begin{figure}[h]
	\includegraphics[scale = 0.22]{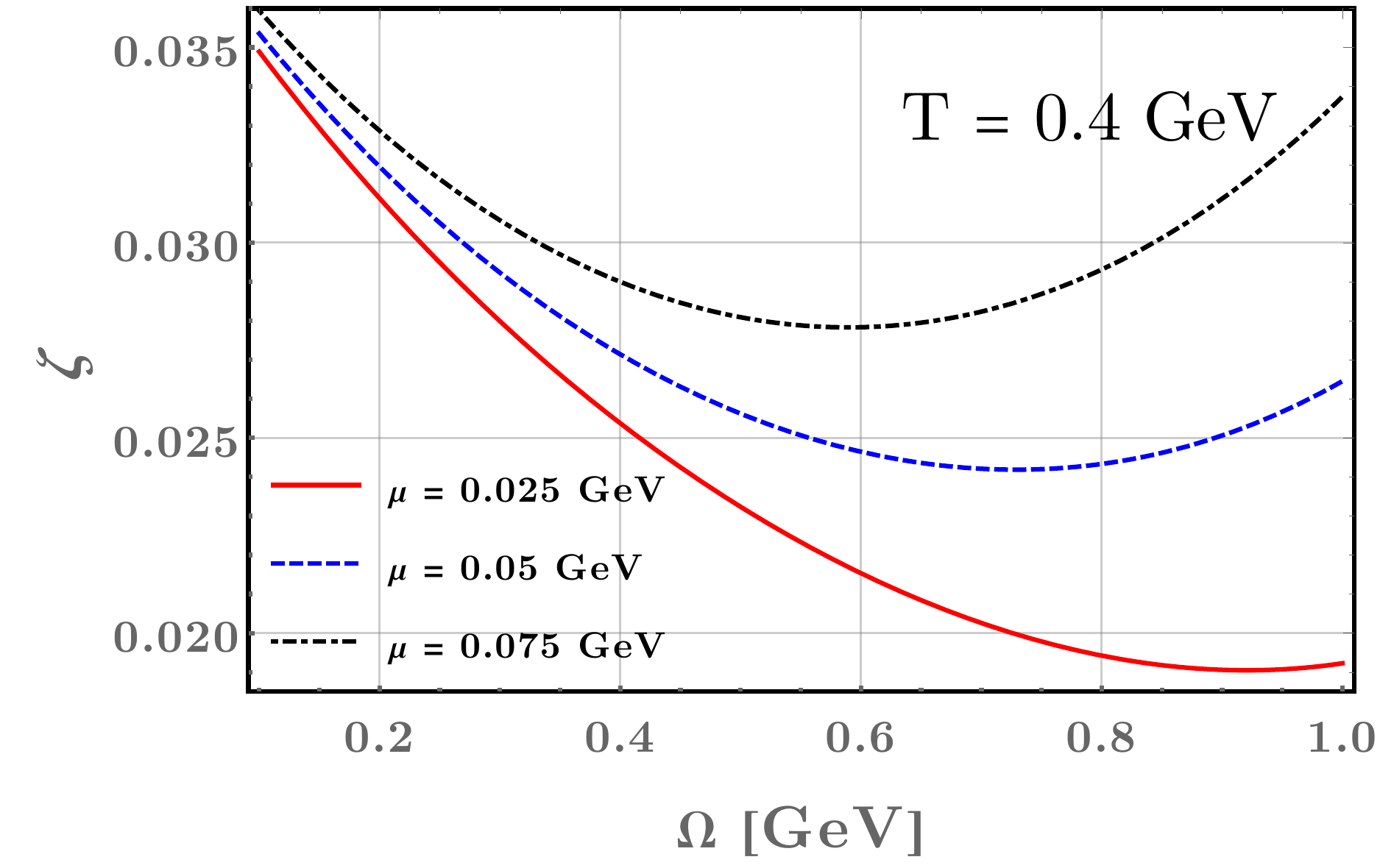}
	\caption{Same as fig. \ref{Zeta_Omg-1}, but for $T =$ 0.4 GeV}
	\label{Zeta_Omg-3}
\end{figure}

Temperatures below the lower bound of the temperature range considered in this work, for a rotating medium constitute two deconfining temperatures as shown by Ref. \cite{Chernodub:2020qah}, due to which the current study might not be able to address the subtleties of the intermediate temperature range, where the deconfining temperature lies. In the deconfined sector, the chemical potential is generally low, unless a degenerate system is considered. Pertaining to this topic, it has been shown by Chernodub {et.al} in 
Ref. \cite{Chernodub:2016kxh} that cold vacuum does not rotate on the imposition of appropiate boundary conditions in the cylinder, since there is no contribution of rotation to the thermodynamic potential. This is an observation that agrees with the conclusions of Ref. \cite{Ebihara:2016fwa}. Therefore the chemical potential considered in this work are $\mu = 0.025$, 0.05 and 0.075 GeV.      

\subsection*{Effect of chemical potential $\mu$ and effective chemical potential $\Omega$}

At low temperatures i.e $T = 0.2 $ GeV, $\zeta$ increases with $\Omega$ as shown by Fig. \ref{Zeta_Omg-1}. As the temperature starts to increase i.e at $T = 0.3$ GeV, as shown by Fig. \ref{Zeta_Omg-2}, the curves show a prominent decreasing behaviour in the region of low $\Omega / T$ values and then the values start to increase as $\Omega / T$ increases. The curves show an increasing trend in an ascending order of $\mu / T$, which is similar to Fig. \ref{Zeta_Omg-1}. Finally in the regime of very high temperature of $T = 0.4$ GeV, shown in Fig. \ref{Zeta_Omg-3}, a behaviour similar to $T = $ 0.3 GeV is observed, but the difference being that the bulk viscosity decreases for most part of the range of $\Omega$ considered, before increasing at large $\Omega$. The common feature of Fig. \ref{Zeta_Omg-1}, \ref{Zeta_Omg-2} and \ref{Zeta_Omg-3} is that the bulk viscosity converges in the limit of low angular frequency which is quite expected. But at the same time we should consider the fact that the range of $\Omega$ considered in this work is very high, suitable for performing calculations with the fermion propagator \cite{Ayala:2021osy} employed in this work  that is valid for high angular velocities of the order of 10$^{-1}$ to 1 GeV. The range of $\Omega$ considered in this work has been experimentally detected in STAR \cite{STAR:2017ckg} experiment. A study of the results in the regime of low angular velocities (i.e below the order of $10^{-1}$ GeV) will require a different expression of the fermion propagator involving an exponential phase factor, which has to be treated for the recovery of translational invariance, before computing the two-point functions of Noether currents and energy-momentum tensors. As discussed before, $\Omega$ enters into the expression of bulk viscosity in two ways viz. (a) distribution function through Matsubara frequency summation (see Appendix \ref{App-A}) and (b) trace evaluation of the gamma matrices of the fermion propagators. Therefore we consider angular velocity to play the role of an effective chemical potential instead of an exact chemical potential. This results in $\zeta$ having different functional dependence on $\mu$ and $\Omega$. The role of $\Omega$ acting as an effective counterpart of $\mu$ has been discussed in Ref. \cite{Chen:2015hfc}. Concerning the behaviour of $\zeta$ in Fig. \ref{Zeta_Omg-1}, \ref{Zeta_Omg-2} and \ref{Zeta_Omg-3}, we will proceed our explanation through a discussion of the dimensionless quantities $\mu / T$ and $\Omega / T$. We should note that there are three energy scales associated with the system i.e $T, \mu$ and $\Omega$. The increase in bulk viscosity with $T$ is a common feature shared by all transport coefficients. At lower temperatures, $\zeta$ as shown by Fig. \ref{Zeta_Omg-1}, both $\mu$ and $\Omega$ contribute to an increase of bulk viscosity. But as $\mu / T$ decreases, which occurs at higher temperatures shown in Fig. \ref{Zeta_Omg-2} and \ref{Zeta_Omg-3}, $\zeta$ decreases upto a particular range of $\Omega$ and then starts to increase with $\Omega$. This can be understood by analysing the Fermi-Dirac distribution function which is of the form $\big[ \exp\big((\pm\mu \pm \Omega)/ T\big) + 1\big]^{-1}$ and $\big[ \exp\big((\mp\mu \pm \Omega)/ T\big) + 1\big]^{-1}$, indicating that $\mu$ and $\Omega$ complement and contrast each other. If the effect of $\mu$ diminishes, then it is compensated or deteoriated by $\Omega$ in the distribution function. The presence of a $\mathcal{O}(\Omega^4)$ dependence in the numerator (\ref{spec-9}) is another functional dependence and this aids to an increase in $\zeta$ as $\Omega$ increases, thus overcoming the exponentially decreasing effect of $\Omega$ in the distribution function. From the observations in Fig. \ref{Zeta_Omg-1}, \ref{Zeta_Omg-2} and \ref{Zeta_Omg-3} it is understood that the effective chemical potential $\Omega$ has a dominant effect at higher values of $\Omega$. Another angle of discussion is concerned with the relaxation time of the spinning fluid. The discussion goes as follows. It is known that bulk viscosity measures the non-uniformity in the expansion or compression of the fluid and is a reflection of the relaxation of the both rotational, vibrational DoFs, and inelastic number changing processes. Since particles posses rotational degrees of freedom, the transport coefficients are associated with the linear and angular momentum contribution, which contribute to the spin polarization effect in the medium of fermions considered here. These effects are responsible for the polarization of the angular momentum caused due to the alignment of angular momentum vectors of rotating particles due to gradients in the medium. Thus, we anticipate that there is a functional dependence of the relaxation time $\tau$ on the angular velocity of the medium. For one-loop Kubo estimation of $\zeta$ considered in this work $\zeta \propto \Gamma^{-1}$, where $\Gamma = \tau^{-1}$. Therefore, an explicit qualitative dependence of $\zeta$ on $\tau(T,\mu,\Omega)$ can be studied by calculating the thermal width for a given interaction.

\section{Conclusions}
\label{Concl}

In this work we have presented a one-loop calculation of bulk viscosity of a system of rotating, hot and dense spin 1/2 fermions, subjected to a very high angular velocity, which for this work has been taken to be between 0.1 - 1.0 GeV. The one-loop calculation serves as a starting point to understand the behaviour of fermions under rotation, before one proceeds to consider the effect of higher loop diagrams and evaluates them via efficient resummation paradigms for transport processes and thermodynamics \cite{Romatschke:2023ztk,Romatschke:2021imm,Weiner:2022kgx,Romatschke:2019ybu,Romatschke:2019wxc,Romatschke:2019rjk,Romatschke:2019gck,Romatschke:2019wxc,Romatschke:2019qbx,Pinto:2020nip,Jeon:1994if,Jeon:1995zm,Lang:2012tt} for a complete pragmatic study. These resummation schemes have been implemented and analysed in the context of large N quantum field theories \cite{Romatschke:2023ztk,Pinto:2020nip,Moshe:2003xn} ( 2+1$d$ O(N) model and 2+1$d$ Gross-Neveu-Yukawa model) which have been able to determine the quantities for arbitrary coupling.
For our work we have considered fermions in a rotating frame studied in tetrad space appropiate for a rotating frame, studied via the Lagrangian (\ref{F-Rot-7}) at a finite chemical potential and angular velocity. The propagator for fermions at large angular velocities has been employed to perform the calculation of the two-point functions of energy-momentum tensors in cylindrical coordinates. The spectral function has been computed using the Saclay method \cite{Laine:2016hma} of the evaluation of the Matsubara frequency sums. It is found that the angular velocity enters the distribution function on computing the Matsubara frequency sums playing the role of an effective chemical potential. There is yet another contribution of the angular velocity in the trace of gamma matrices which carry carry a $\Omega$ dependence and from the temporal component of the covariant derivative. The presence of the angular velocity in the distribution functions and the traces prompts us to label it as an effective chemical potential instead of an exact one. The bulk viscosity as a function of temperature, chemical potential and angular velocity has been calculated by using the Kubo formula (\ref{spec-1}) for bulk viscosity and the corresponding plots as a function of angular velocity, for different values of temperature and chemical potentials are plotted. The results show that bulk viscosity increases with temperature (in the range of purely deconfined state), whereas the variation with chemical potential and angular velocity is non-trivial. It is found that the chemical potential and angular velocity complement each other, with the angular velocity becoming the dominant energy scale as it increases. The results presented in this work are valid for very large angular velocities.

\section*{Acknowledgements}
I am indebted to Maxim Chernodub for his deep insights, suggestions and discussions on the problem. I thank Amaresh Jaiswal for his encouragement and advice to work on this topic. I am grateful to my colleagues Pushpa Panday, Abhishek Tiwari, Salman Ahamad Khan, Sumit and Debarshi Dey for the valuable discussions. I thank Rajeev Singh for his useful comments and suggestions on the manuscript.  This work has been funded by the Institute Post Doctoral Scheme of Indian Institute of Technology Roorkee under the grant IITR/Estt-(A)-Rect-Cell-E-5001(130)18490.

\appendix

\section{Matsubara frequency sums at finite  $T$, $\mu$ and $\Omega$}
\label{App-A}
In this work we have employed Matsubara frequency summation \cite{Laine:2016hma} to calculate the bulk viscosity. Here we provide the details of the Matsubara frequency summation used for calculating the spectral function of energy-momentum tensors. At finite temperature

$$p_0 \to i\widetilde{\omega}_N = (2N + 1)\pi T,~q_0 \to i\nu_N = 2\pi NT,~\int\frac{d^4p}{(2\pi)^4}\equiv \sumintof = iT\sum_{N = -\infty}^{+\infty}\int\frac{d^3p}{(2\pi)^3} ,\text{~~where~~} N\in \mathbb{Z} $$ 

where 
$p_0$ and $q_0$ are the temporal components of the 4-momentum of the fermion and boson in the one-loop diagram. The fermion propagator under rotation is given by Eq. (\ref{F-Rot-8}) which reads as follows

\bea
S(p) &=& \frac{\big( p_0 + \frac{\Omega}{2} - p_z + ip_\perp   \big)\big(\gamma_0 + \gamma_3\big) + m\big(1 + \gamma_5\big)}{\big(p_0 + \frac{\Omega}{2}\big)^2 - \vec{p}^2 - m^2 + i\epsilon}\mathcal{O}^+ + \frac{\big( p_0 - \frac{\Omega}{2} + p_z - ip_\perp   \big)\big(\gamma_0 - \gamma_3\big) + m\big(1 + \gamma_5\big)}{\big(p_0 - \frac{\Omega}{2}\big)^2 - \vec{p}^2 - m^2 + i\epsilon}\mathcal{O}^-~.\nn\\
\label{App-A-1}
\eea 

In one-loop Kubo calculations we encounter fermionic Matsubara frequency summations of the type
\begin{enumerate} 
	\item\bea
	T\sum_{\{p_N\}}\frac{1}{\big[ i\widetilde{\omega}_N + \frac{\Omega}{2} + \mu  \big]^2 - \vec{p}^2 - m^2}    \frac{1}{\big[ i\nu_N - i\widetilde{\omega}_N + \frac{\Omega}{2} - \mu  \big]^2 - \vec{k}^2 - m^2}
	\label{App-A-2}
	\eea 
	
	\item\bea
	T\sum_{\{p_N\}}\frac{1}{\big[ i\widetilde{\omega}_N + \frac{\Omega}{2} + \mu  \big]^2 - \vec{p}^2 - m^2}    \frac{1}{\big[ i\nu_N - i\widetilde{\omega}_N - \frac{\Omega}{2} - \mu  \big]^2 - \vec{k}^2 - m^2}
	\label{App-A-3}
	\eea 
	
	\item\bea
	T\sum_{\{p_N\}}\frac{1}{\big[ i\widetilde{\omega}_N - \frac{\Omega}{2} + \mu  \big]^2 - \vec{p}^2 - m^2}    \frac{1}{\big[ i\nu_N - i\widetilde{\omega}_N + \frac{\Omega}{2} - \mu  \big]^2 - \vec{k}^2 - m^2}
	\label{App-A-4}
	\eea

	\item\bea
	T\sum_{\{p_N\}}\frac{1}{\big[ i\widetilde{\omega}_N - \frac{\Omega}{2} + \mu  \big]^2 - \vec{p}^2 - m^2}    \frac{1}{\big[ i\nu_N - i\widetilde{\omega}_N - \frac{\Omega}{2} - \mu  \big]^2 - \vec{k}^2 - m^2}
	\label{App-A-5}
	\eea 
\end{enumerate}
Following the Saclay method \cite{Laine:2016hma} of the evaluation of Matsubara frequency sums we obtain the results for the above frequency summations in Eqs.(\ref{App-A-2})-(\ref{App-A-5}) in Eqs.(\ref{App-A-6})-(\ref{App-A-9}) as follows : 

	\bea
	(a)&&T\sum_{\{p_N\}}\frac{1}{\big[ i\widetilde{\omega}_N + \frac{\Omega}{2} + \mu  \big]^2 - \vec{p}^2 - m^2}    \frac{1}{\big[ i\nu_N - i\widetilde{\omega}_N + \frac{\Omega}{2} - \mu  \big]^2 - \vec{k}^2 - m^2} \nn \\
	&=& \frac{1}{4E_pE_k}\Bigg\{  \frac{n_F(E_p + \mu + \Omega/2) + n_F(E_k - \mu + \Omega/2) - 1}{q_0 - E_p - E_k} + \frac{n_F(E_k + \mu + \Omega/2) - n_F(E_p + \mu + \Omega/2)}{q_0 + E_k - E_p}  \nn \\
	&+& \frac{n_F(E_p - \mu - \Omega/2) - n_F(E_k - \mu - \Omega/2)}{q_0 + E_p - E_k} + \frac{1 - n_F(E_p - \mu - \Omega/2) - n_F(E_k + \mu - \Omega/2)}{q_0 + E_k + E_p} \Bigg\}\nn \\
	\label{App-A-6}
	\eea

	\bea
	(b)&&T\sum_{\{p_N\}}\frac{1}{\big[ i\widetilde{\omega}_N + \frac{\Omega}{2} + \mu  \big]^2 - \vec{p}^2 - m^2}    \frac{1}{\big[ i\nu_N - i\widetilde{\omega}_N - \frac{\Omega}{2} - \mu  \big]^2 - \vec{k}^2 - m^2} \nn \\
	&=& \frac{1}{4E_pE_k}\Bigg\{  \frac{n_F(E_p + \mu + \Omega/2) + n_F(E_k - \mu - \Omega/2) - 1}{q_0 - E_p - E_k} + \frac{n_F(E_k + \mu + \Omega/2) - n_F(E_p + \mu + \Omega/2)}{q_0 + E_k - E_p}  \nn \\
	&+& \frac{n_F(E_p - \mu - \Omega/2) - n_F(E_k - \mu - \Omega/2)}{q_0 + E_p - E_k} + \frac{1 - n_F(E_p - \mu - \Omega/2) - n_F(E_k + \mu + \Omega/2)}{q_0 + E_k + E_p} \Bigg\}\nn \\
	\label{App-A-7}
	\eea 
	
    \bea
	(c)&&T\sum_{\{p_N\}}\frac{1}{\big[ i\widetilde{\omega}_N - \frac{\Omega}{2} + \mu  \big]^2 - \vec{p}^2 - m^2}    \frac{1}{\big[ i\nu_N - i\widetilde{\omega}_N + \frac{\Omega}{2} - \mu  \big]^2 - \vec{k}^2 - m^2} \nn \\
	&=& \frac{1}{4E_pE_k}\Bigg\{  \frac{n_F(E_p + \mu - \Omega/2) + n_F(E_k - \mu + \Omega/2) - 1}{q_0 - E_p - E_k} + \frac{n_F(E_k + \mu - \Omega/2) - n_F(E_k + \mu - \Omega/2)}{q_0 + E_k - E_p}  \nn \\
	&+& \frac{n_F(E_p - \mu + \Omega/2) - n_F(E_k - \mu + \Omega/2)}{q_0 + E_p - E_k} + \frac{1 - n_F(E_p - \mu + \Omega/2) - n_F(E_k + \mu - \Omega/2)}{q_0 + E_k + E_p} \Bigg\}\nn \\
	\label{App-A-8}
	\eea
	
	\bea
	(d)&&T\sum_{\{p_N\}}\frac{1}{\big[ i\widetilde{\omega}_N - \frac{\Omega}{2} + \mu  \big]^2 - \vec{p}^2 - m^2}    \frac{1}{\big[ i\nu_N - i\widetilde{\omega}_N - \frac{\Omega}{2} - \mu  \big]^2 - \vec{k}^2 - m^2} \nn \\
	&=& \frac{1}{4E_pE_k}\Bigg\{  \frac{n_F(E_p + \mu - \Omega/2) + n_F(E_k - \mu - \Omega/2) - 1}{q_0 - E_p - E_k} + \frac{n_F(E_k + \mu - \Omega/2) - n_F(E_k + \mu - \Omega/2)}{q_0 + E_k - E_p}  \nn \\
	&+& \frac{n_F(E_p - \mu + \Omega/2) - n_F(E_k - \mu + \Omega/2)}{q_0 + E_p - E_k} + \frac{1 - n_F(E_p - \mu + \Omega/2) - n_F(E_k + \mu - \Omega/2)}{q_0 + E_k + E_p} \Bigg\}\nn \\
	\label{App-A-9}
	\eea



\end{document}